\documentclass[reprint, superscriptaddress, amsmath, amssymb, aps, pra]{revtex4-2}
\usepackage{graphicx}
\usepackage{dcolumn} 
\usepackage{xcolor}
\usepackage[absolute]{textpos}
\usepackage{bm}
\usepackage{braket}
\usepackage{mathtools}
\usepackage{verbatim}

\usepackage{amsmath,amssymb,bbm,bm}
\usepackage{braket}
\usepackage[pdftex,colorlinks=true,bookmarks=false,citecolor=blue,urlcolor=blue]{hyperref}
\usepackage[absolute]{textpos}
\usepackage{xcolor}
\definecolor{ASUblue}{RGB}{0,163,224}
\definecolor{darkblue}{rgb}{0,0.3,0.7}

\DeclareMathOperator{\Tr}{Tr}
\DeclareMathOperator*{\argmax}{arg\,max}
\newcommand\bx{\mathbf{x}}
\newcommand\cD{{\mathcal{D}}}
\newcommand\bcD{{\bm{\mathcal{D}}}}

\newcommand\cN{\mathcal{N}}

\newcommand\cF{\mathcal{F}}
\newcommand\cO{\mathcal{O}}
\newcommand\imag{\mathrm{i}}
\newcommand\kvac{\ket{\text{vac}}}

\DeclareMathOperator{\Real}{\mathfrak{Re}}
\newcommand\hpi{\hat{\pi}}
\newcommand\hrho{\hat{\rho}}
\newcommand\hcF{\hat{\mathcal{F}}}

\definecolor{teal}{RGB}{26,157,150}
\definecolor{peach}{HTML}{DE7E5D}

\hypersetup{
colorlinks=true,
linkcolor=darkblue,
filecolor=blue,
citecolor=darkblue,  
urlcolor=darkblue,
}

\usepackage[capitalize]{cleveref}
\crefname{section}{Sec.}{Secs.}
\Crefname{section}{Section}{Sections}

\crefrangelabelformat{section}{#3#1#4--#5\crefstripprefix{#1}{#2}#6}
\crefrangelabelformat{figure}{#3#1#4--#5\crefstripprefix{#1}{#2}#6}
\crefrangelabelformat{equation}{(#3#1#4--#5\crefstripprefix{#1}{#2}#6)}

\crefmultiformat{equation}%
{\edef\crefstripprefixinfo{#1}Eqs.~(#2#1#3}%
{,#2\crefstripprefix{\crefstripprefixinfo}{#1}#3)}%
{,#2\crefstripprefix{\crefstripprefixinfo}{#1}#3}%
{,#2\crefstripprefix{\crefstripprefixinfo}{#1}#3)}

\begin{document}

\title{Opportunistic full reconstruction of 100-dimensional frequency-bin quantum states}

\author{Hanson~H.~Nguyen}
\email{hansonng@berkeley.edu}
\affiliation{Department of Electrical Engineering and Computer Sciences, University of California, Berkeley, Berkeley, California 94720, USA}
\affiliation{Elmore Family School of Electrical and Computer Engineering and Purdue Quantum Science and Engineering Institute, Purdue University, West Lafayette, Indiana 47907, USA}
\author{Zachary Goisman}
\affiliation{Elmore Family School of Electrical and Computer Engineering and Purdue Quantum Science and Engineering Institute, Purdue University, West Lafayette, Indiana 47907, USA}
\author{Chen-You Su}
\affiliation{Elmore Family School of Electrical and Computer Engineering and Purdue Quantum Science and Engineering Institute, Purdue University, West Lafayette, Indiana 47907, USA}
\affiliation{Department of Physics and Astronomy, Purdue University, West Lafayette, Indiana 47907, USA}

\author{Diego Maragnano}
\affiliation{Dipartimento di Fisica ``A. Volta'', Università di Pavia, Via Bassi 6, 27100 Pavia, Italy}

\author{Marco Liscidini}
\affiliation{Dipartimento di Fisica ``A. Volta'', Università di Pavia, Via Bassi 6, 27100 Pavia, Italy}

\author{Jason~D.~McKinney}
\affiliation{Elmore Family School of Electrical and Computer Engineering and Purdue Quantum Science and Engineering Institute, Purdue University, West Lafayette, Indiana 47907, USA}

\author{Hsuan-Hao Lu}
\affiliation{Elmore Family School of Electrical and Computer Engineering and Purdue Quantum Science and Engineering Institute, Purdue University, West Lafayette, Indiana 47907, USA}

\author{Joseph~M.~Lukens}
\email{jlukens@purdue.edu}
\affiliation{Elmore Family School of Electrical and Computer Engineering and Purdue Quantum Science and Engineering Institute, Purdue University, West Lafayette, Indiana 47907, USA}
\affiliation{Quantum Information Science Section, Oak Ridge National Laboratory, Oak Ridge, Tennessee 37831, USA}

\begin{abstract}
Frequency-bin encoding provides a scalable platform for photonic quantum information processing by enabling high-dimensional qudit states compatible with the telecommunications infrastructure. However, characterizing these systems remains challenging, as conventional quantum state tomography requires measurement resources that increase rapidly with dimensionality, compounded by the difficulty of implementing controllable measurements across many frequency modes at once. Here, we address these challenges and demonstrate threshold quantum state tomography of entangled frequency-bin qudits using adaptive measurements that exploit the sparsity of highly entangled biphoton frequency combs. 
Combined with Bayesian inference, we experimentally demonstrate full state estimation of biphoton frequency combs with Hilbert space dimensions up to $d^2=100$---a 56\% increase over the previous record---all while performing a mere 1.9\% of the measurements required by conventional tomographically complete approaches.
This work confirms the promise of measurement-aware approaches for characterizing high-dimensional photonic quantum systems, unlocking new tools for full tomography that are hardware-efficient, state-aware, and free from \emph{a priori} assumptions about state structure.
\end{abstract}

\maketitle

\section{Introduction}
Frequency-bin encoding \cite{Lu2023} 
has become an increasingly important platform for quantum communications and networking~\cite{Khodadad2025, Tagliavacche2025} due to its inherent compatibility with the telecommunications infrastructure and photonic integrated circuits~\cite{Kues2017, Clementi2023, Myilswamy2025, Lukens2026}. 
In the frequency domain, unique advantages arise from the large number of entangled spectral bins accessible in biphoton frequency combs (BFCs), with experimental demonstrations progressively increasing the number of accessible frequency modes to the thousands in standard systems~\cite{Ikuta2019, Chang2021, Yamazaki2022, Chang2024}.
However, these same features that make frequency-bin systems attractive for quantum technologies also present significant challenges for characterization and verification.
Performing superposition measurements in the frequency-bin basis has remained a persistent experimental challenge because distinct frequency modes do not naturally couple under passive linear optics, requiring sophisticated active optical techniques to coherently mix multiple bins \cite{Clemmen2016, Kues2017, Lu2018, lu2019controlled, Oliver2025}. These frequency-specific challenges compound with the exponential scaling of quantum state tomography (QST) in general, for any encoding: an arbitrary $n$-qudit state $\rho$ requires the estimation of $d^{2n}-1$ independent real parameters, subject to the constraints $\rho \succeq 0$ and $\Tr\rho = 1$ \cite{James2001,  Lvovsky2004, Cramer2010, Nielsen2012, Smolin2012}. Consequently, as the dimensionality of frequency-bin quantum systems increases, so do the measurement and computational resources required for QST.

At a high level, these issues can be divided into experimental and computational challenges---the former concerned with the physical implementation of desired quantum measurements \cite{deBurgh2008, Adamson2010, Huszr2012}, the latter with the optimal selection of these settings and the subsequent reconstruction of the quantum state or its properties from the collected data~\cite{blume2010optimal, Gross2010, Cramer2010}. These challenges are distinct, yet interdependent: hardware limitations can inform the algorithmic frameworks available for state reconstruction, while algorithmic flexibility can extract useful information from selective measurements \cite{Lukens_2020, Huang2020, Lukens2021}.

For instance, current state-of-the-art high-dimensional frequency-bin QST combines a single electro-optic phase modulator (EOPM) with randomly selected modulation indices \cite{Elben2023} and Bayesian inference \cite{Lukens_2020} to accurately reconstruct entangled frequency-bin qudits up to $d^2 = 64$ from tomographically incomplete measurements \cite{Lu2022b} with low uncertainty. Despite this experimental simplicity, the approach faces significant scaling limitations. Experimentally, the reliance on $d^2$-outcome measurements introduces quadratic scaling in the required settings per basis under two-detector raster scans. Computationally, estimating $\mathcal{O}(d^{4})$ unknown parameters causes Bayesian inference Markov chains to grow exponentially in time, requiring 161~h of wall-clock time for the $d=8$ case in Ref.~\cite{Lu2022b}.

Recently, however, two new discoveries have become poised to break this stagnation. Computationally, unorthodox parallel Markov Chain Monte Carlo (MCMC) strategies have opened a path for high-performance computing in Bayesian QST, already leading to empirical 100-fold speedups in four-qubit experiments \cite{Nguyen2025}. Experimentally, threshold quantum state tomography (tQST) has been proposed as an alternative measurement protocol that exploits the Cauchy--Schwarz inequality to selectively characterize only the significant off-diagonal elements of a density matrix \cite{Maragnano2023, Binosi2024, Binosi2024b, Garberoglio2025, Caruccio2025}. 
Auspiciously, tQST is uniquely well-positioned for probing BFCs: (i) due to frequency entanglement, these biphoton states are naturally sparse in the computational basis; and (ii) the experimental advantages of tQST are most pronounced when expressed as two-outcome projections, which correspond directly to the number of experimental settings---and hence time---in two-detector frequency-bin qudit experiments. 

In this work, we implement tQST of frequency-bin quantum states for the first time. Leveraging electro-optic phase modulation and pulse shaping for targeted frequency-bin interference, our technique additionally incorporates parallelized Bayesian inference in post-processing, quantifying uncertainty that automatically reflects both statistical noise and informational incompleteness. 
In the process, we realize full frequency-bin tomography of entangled frequency-bin qudits up to $d^2=100$-dimensional Hilbert spaces, achieving fidelities $\cF>0.845$ with respect to the hypothesized ground truth. Because the number of projective measurements scales as $2d^2-d$---a square-root reduction compared to conventional tomography---estimates are obtained with only 1.9\% of the projections required in standard tomography for the record-high 100-dimensional cases. Overall, our approach unlocks new diagnostic tools for frequency-bin quantum information processing, revealing state properties in a measurement-efficient, yet assumption-free, fashion.

\section{Theory}
\label{sec:freqBinTqst}
\begin{figure}[t!]
  \centering
  \includegraphics[width=0.5\textwidth]{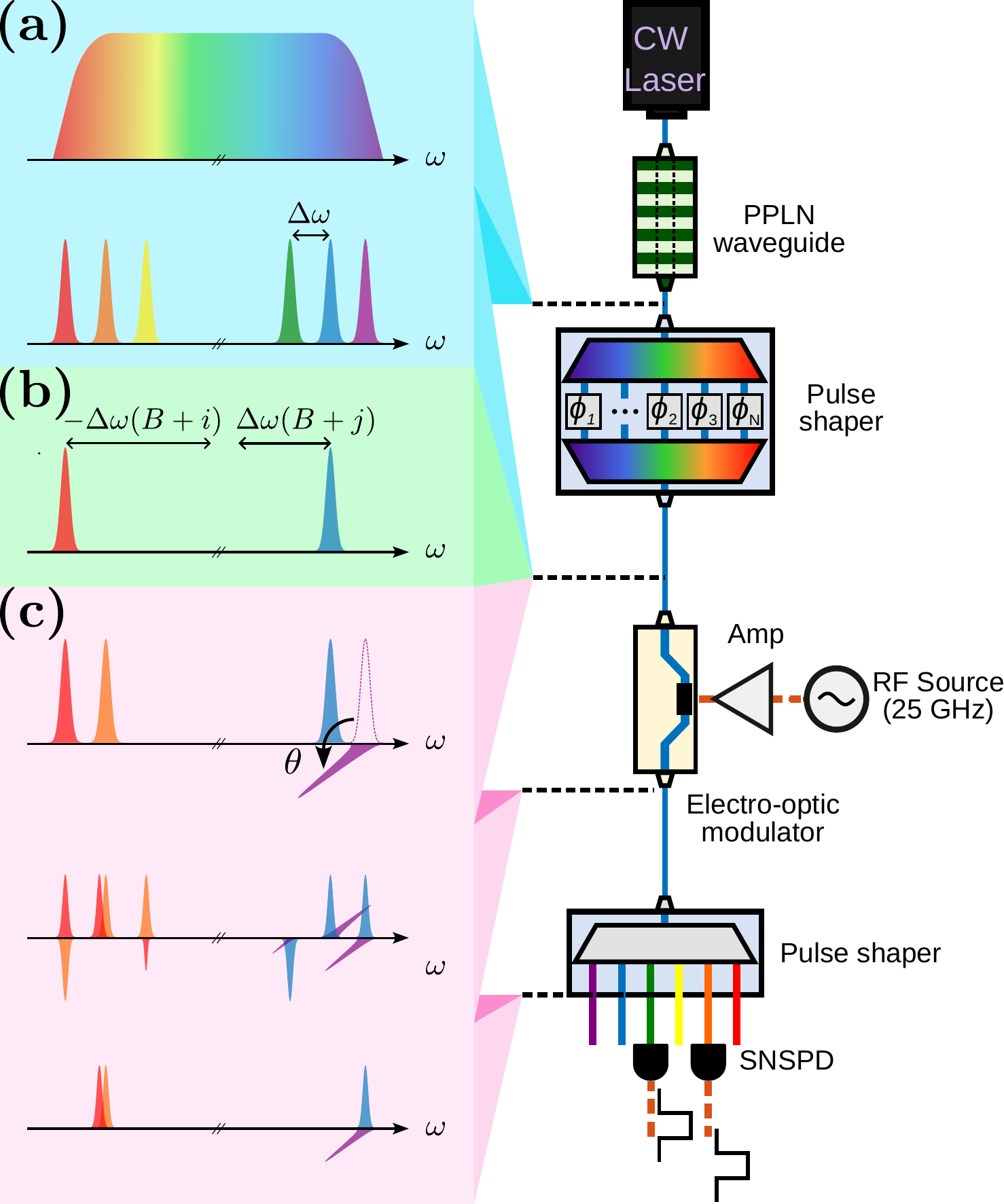}
\caption{Experimental setup illustrating the state-preparation and measurement stages. (a) Generation of a broadband frequency-entangled state via spontaneous parametric down-conversion in a periodically poled lithium niobate waveguide pumped by a 775~nm continuous-wave laser, followed by definition of the frequency bins with a Fourier-transform pulse shaper. (b) Measurement of computational-basis correlations to characterize the sparsity and determine the measurements used in threshold quantum state tomography. (c) Measurement of targeted off-diagonal coherences through selection of frequency bin pairs (with phase shift $\theta$ applied to one bin), frequency mixing using an electro-optic phase modulator, and coincidence measurements between bins $\omega_{-i}$ and $\omega_j$.}
\label{fig1}
\end{figure}

\Cref{fig1} illustrates the full tQST scheme for characterizing  BFCs. The hardware configuration closely resembles previous work in high-dimensional frequency-bin tomography~\cite{Lu2022b} and rank certification~\cite{Cabrejo2023}. 
We consider the case in which the ideal state is a high-dimensional maximally entangled state of the form $\rho_0 = \frac{1}{d}\sum_{i,j\in[d]} \ket{ii}\bra{jj}$, where $[d]\equiv\{1,...,d\}$ and the computational basis states $\ket{ii} = a_{-i}^\dagger a_{i}^\dagger \ket{\mathrm{vac}}$ correspond to photon pairs centered at frequencies $\omega_{\mp i} = \omega_0 \mp (B + i)\Omega$; $B$ is a guardband specifically introduced to suppress spectral crosstalk between the signal and idler bands during the characterization procedure---a  complication that can be avoided if the signal and idler bins are first demultiplexed into separate paths. Here, we choose $B = 5$, sufficiently large for all modulation indices considered.

In general, the frequency-bin density matrix lives in a $d^2$-dimensional Hilbert space and therefore contains $d^4$ total elements, described formally in terms of matrix elements $\rho_{(ij)(k\ell)}=\braket{ij|\rho|k\ell}$ as 
\begin{equation}
\label{eq:rho}
\rho = \sum_{i,j,k,\ell\in[d]} \rho_{(ij)(k\ell)}\ket{ij}\bra{k\ell}.
\end{equation}
By the Cauchy-Schwarz inequality, $|\rho_{(ij)(k\ell)}|\leq\sqrt{\rho_{(ij)(ij)}\rho_{(k\ell)(k\ell)}}$, which in isolation is merely an implication of positive semidefiniteness ($\rho\succeq0$). However, to the extent that $\rho\approx\rho_0$ and hence $\rho_{(ij)(ij)}\approx\frac{1}{d}\delta_{ij}$, it significantly constrains magnitudes of the off-diagonal elements $\rho_{(ij)(k\ell)}$ for mismatched  $i\neq j$ and $k\neq\ell$.

tQST leverages this recognition by first measuring the diagonal elements $\rho_{(ij)(ij)}$,  defining a threshold $t$, and finally performing measurements only of observables relevant to determine the coherences $\rho_{(ij)(k\ell)}$ whose Cauchy-Schwarz ceilings exceed a chosen threshold $t$: $\sqrt{\rho_{(ij)(ij)}\rho_{(k\ell)(k\ell)}}>t$~\cite{Binosi2024}. In the context of highly correlated BFCs for which $\rho_{(ij)(ij)}<\rho_{(ii)(ii)} \,\forall\,j\neq i$, it is natural to define the threshold such that only coherences  between spectrally correlated pairs are considered:
\begin{equation}
\label{eq:threshold}
\sqrt{\rho_{(ij)(ij)}\rho_{(kk)(kk)}} < t < \sqrt{\rho_{(ii)(ii)}\rho_{(jj)(jj)}}
\end{equation}
for all $i,j,k\in[d]$ such that $i\neq j$. With this choice, only off-diagonal elements of the form $\rho_{(ii)(jj)}$ are targeted in tQST superposition measurements: 
$\frac{1}{2}d(d-1)$ elements of the total number of $\frac{1}{2}d^2(d^2-1)$ off-diagonals for a generic state (considering only the upper diagonals due to Hermiticity).

Before proceeding, it is important to note two important characteristics of tQST that distinguish it from techniques for accelerated tomography based on structural assumptions of the state itself~\cite{Qin2026}, such as low rankness~\cite{Gross2010}, matrix product operator form~\cite{Cramer2010}, or restricted Boltzmann machine representations~\cite{Torlai2018}. First, tQST does not make any assumptions on the state to be characterized; therefore, the specific nature of the correlations need not be as uniform and straightforward as in the present BFC to achieve significant advantages with respect to most other approaches.
Second, in tQST the selection of $t$ determines the number of measurements to be performed following the ranking of the most relevant off-diagonal terms of the density matrix according to the Cauchy-Schwarz inequality. The choice of $t$ is not unique; for instance, it can be chosen objectively based on the uniformity of the diagonal of the density matrix, as quantified by the Gini index~\cite{Caruccio2025}, or chosen depending on the available resources. Importantly, such a choice does not place any restrictions on unprobed density matrix elements. In the present case, the mismatched components $\rho_{(ij)(k\ell)}$ with $j\neq i$ and $\ell\neq k$ are free to assume any values consistent with the measurement outcomes and physical constraints.
Thus, tQST is assumption-free: it exploits observed sparsity without imposing sparsity explicitly. Conveniently, such a condition is aligned with the philosophy behind Bayesian inference, which is designed to return principled estimators for any set of experimental observations, with the posterior uncertainty automatically accounting for any missing information~\cite{Robert1999, MacKay2003, blume2010optimal,Lukens_2020,Nguyen2025}.

To obtain a full characterization of the BFC state in tQST, we must therefore determine both the magnitude and phase of the off-diagonal elements $\rho_{(ii)(jj)}$ identified in the thresholding procedure. For concreteness, we take $i < j$ in all that follows such that bins $\omega_{\pm j}$ are always further from the BFC center than bins $\omega_{\pm i}$. 
We consider projection onto the state $\ket{u_{ij}v_{ij}}\equiv\ket{u_{ij}}\otimes\ket{v_{ij}}$, where
\begin{equation}
\label{eq:projStates}
\begin{split}
\ket{u_{ij}} & = \frac{1}{\sqrt{2}}(a_{-i}^\dagger+a_{-j}^\dagger)\kvac \\ \ket{v_{ij}} & =\frac{1}{\sqrt{2}}(a_i^\dagger+e^{-\imag\theta}a_j^\dagger)\kvac
\end{split}
\end{equation}
denote two-bin superposition states of the idler and signal, respectively.
The probability  of obtaining this outcome, $p_{ij}=\braket{u_{ij}v_{ij}|\rho|u_{ij}v_{ij}}$,  is then given in terms of the density matrix elements by
\begin{widetext}
\begin{multline}
\label{eq:probIdeal}
p_{ij} = 
\frac{1}{4}\bigg\{\rho_{(ii)(ii)} + \rho_{(jj)(jj)} + 2\Real e^{-\imag\theta}\rho_{(ii)(jj)} \\
+\rho_{(ij)(ij)} + \rho_{(ji)(ji)} + 
2\Real\left[\rho_{(ii)(ji)}+\rho_{(ij)(jj)}+e^{-\imag\theta}\rho_{(ii)(ij)}+e^{-\imag\theta}\rho_{(ji)(ij)}+e^{-\imag\theta}\rho_{(ji)(jj)} \right]  \bigg\}.
\end{multline}
\end{widetext}
Because all the diagonal density matrix elements are assumed to have been probed and all off-diagonal elements in the second row are upper-bounded by $t$, $\rho_{(ii)(jj)}$ represents the only unknown quantity in \cref{eq:probIdeal} with magnitude possibly greater than $t$. Hence, by considering $\theta\in\{0,\frac{\pi}{2}\}$, both the real and imaginary parts of $\rho_{(ii)(jj)}$ can be extracted, up to additional terms that---although not neglected---are nevertheless bounded.

To simplify the experimental implementation,  we do not synthesize the exact frequency-bin projections in \cref{eq:projStates} directly---which \emph{could} be realized by a three-element quantum frequency processor with variable radio frequency (RF)~\cite{Lu2018,Lingaraju2022}---but rather apply line-by-line pulse shaping followed by a single layer of electro-optic phase modulation driven by a single-frequency RF tone at $\Omega$ with tunable modulation amplitude $\beta$, as outlined in \cref{fig1}. Defining the discrete spectral filter $h_n$ for bin $\omega_n$ and assuming  phase modulation $e^{-\imag\beta\sin\Omega t}$, the output annihilation operators for the idler in bin $\omega_{-i}$ and the signal in bin $\omega_j$ can be written as~\cite{Lu2022b}
\begin{equation}
\begin{split}
\label{eq:annihilation}
b_{-i} &= \sum_{k=1}^d J_{-i+k}(\beta)h_{-k}a_{-k} \\
b_j &= \sum_{\ell=1}^d J_{j-\ell}(\beta)h_\ell a_\ell.
\end{split}
\end{equation}

By passing only the four frequency bins $\omega_{\pm i}$ and $\omega_{\pm j}$ and applying a phase shift $\theta$ on $\omega_j$ via the filter 
$h_n=\delta[n+i]+\delta[n+j]+\delta[n-i]+e^{\imag\theta}\delta[n-j]$, a coincidence  between bins $\omega_{-i}$ and $\omega_j$ occurs with probability
\begin{widetext}
\begin{equation}
\begin{split}
\label{eq:probBessel}
p_{ij} & = \braket{\text{vac}| b_{-i}b_j \rho b_{-i}^\dagger b_j^\dagger|\text{vac}} \\
& = \sum_{k,\ell,k',\ell'\in[d]}
\rho_{(k\ell)(k'\ell')}
\braket{\text{vac}|b_{-i}b_j|k\ell}
\braket{k'\ell'|b_{-i}^\dagger b_j^\dagger|\text{vac}}, \\
& = J_0^2(\beta)J_{j-i}^2(\beta)\left[\rho_{(ii)(ii)}+\rho_{(jj)(jj)} + 2\Real e^{-\imag\theta}\rho_{(ii)(jj)} \right] + \cO(t),
\end{split}
\end{equation}
\end{widetext}
where $\cO(t)$ denotes terms comprising density matrix elements bounded by the threshold $t$. 
Significantly, the bracketed terms in \cref{eq:probBessel} match the first row of \cref{eq:probIdeal} except for scaling by a product of Bessel functions, thus isolating the real and imaginary parts of $\rho_{(ii)(jj)}$ as designed by tQST. Since, for all $\beta\in\mathbb{R}$ and $j\neq i$, $J_0^2(\beta)J_{j-i}^2(\beta)<\frac{1}{4}$ holds, our electro-optic approach has a limited efficiency even in the lossless case. Yet in the context of QST---where there is no need for further downstream state processing---this efficiency reduction merely lowers the detection rate with no additional errors.

As an aside, we note that \cref{eq:annihilation} assumes the \emph{same} modulation index $\beta$ for both the idler and signal projections, a choice motivated by the symmetry of the BFC. The off-diagonal elements of interest $\rho_{(ii)(jj)}$ correspond to superpositions with the same frequency separation for the two photons: $\omega_{-i}-\omega_{-j}=\omega_j-\omega_i = (j-i)\Omega$. Consequently, a modulation index optimized for efficiently mixing idler bins will optimize the signal bin mixing as well. Experimentally, such symmetry allows us to consider a single EOPM in \cref{fig1} with no compromise to tQST performance. Nonetheless, it would be straightforward to generalize to separate modulators in cases where the coherences lack this symmetry (or when the photons are physically separated), by drawing on configurations from prior work in nonlocal electro-optic phase modulation~\cite{Harris2008, Sensarn2009, Seshadri2022b, Chapman2025}.

\section{Results}
\label{sec:results}

\begin{figure}[tb!]
  \centering
  \includegraphics[width=0.5\textwidth]{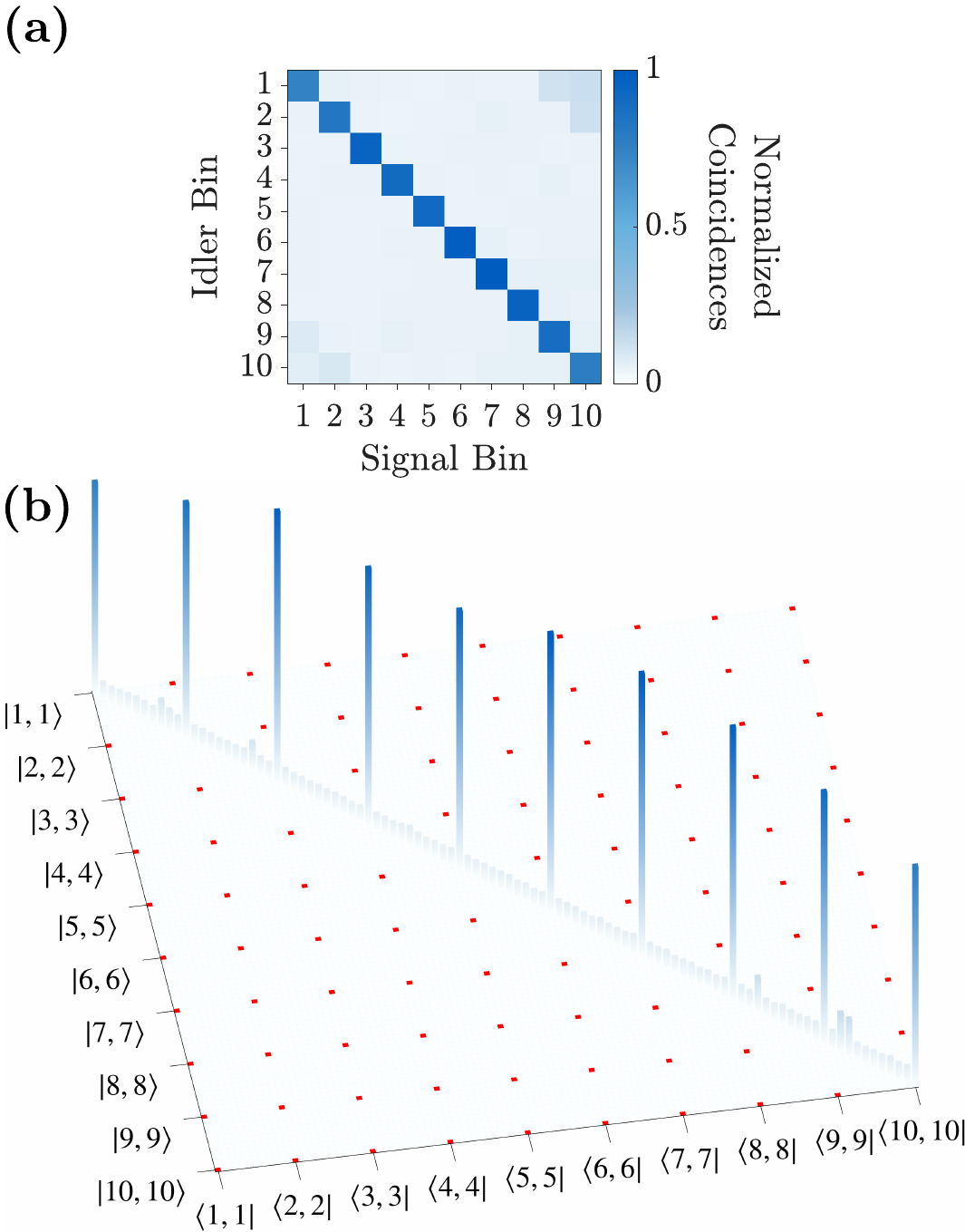}
\caption{Experimental computational-basis measurements. (a) Joint spectral intensity of $d=10$ frequency bins per photon, exhibiting a coincidence-to-accidental ratio of 31.  (b) Corresponding diagonal density matrix values (blue) and off-diagonal elements identified by thresholding to be measured (red).}
\label{fig2}
\end{figure}

In the experimental setup shown in Fig.~\ref{fig1}, we pump a type-0 periodically poled lithium niobate (PPLN) waveguide mixer (HC Photonics) with a 775~nm continuous-wave laser to produce broadband entangled photons through spontaneous parametric downconversion. We then carve a BFC from the output using a Fourier-transform pulse shaper (Waveshaper 4000A), where each bin is 20~GHz wide and spaced by 25~GHz. The same pulse shaper is used to select specific bins for measurement, which are then mixed by a thin-film lithium-niobate electro-optic phase modulator (EOPM; Agiltron) driven by an RF signal generator. A second pulse shaper is employed as a wavelength-selective switch (WSS), routing the output to two superconducting nanowire single-photon detectors (Quantum Opus) connected to a time tagger (Swabian) for coincidence analysis.  

Sweeping across pairs of signal-idler frequency bins with RF drive off, we obtain the joint spectral intensity (JSI) in \cref{fig2}(a) for the $d=10$ case at 1~s integration time per point, which shows appreciable weights only for matched frequency bins $\omega_{\pm i}$, with a coincidence-to-accidental ratio (CAR) of $\sim$31. Visually, we can unfold the results of this JSI into the diagonal of a tentative density matrix~\cite{footnoteEmpiricalFreq} as depicted in \cref{fig2}(b), from which the thresholding regime is evident: namely, for any $t>0.072$ 
only the off-diagonal terms $\rho_{(ii)(jj)}$ [red dots in \cref{fig2}(b)] 
remain in the tQST measurement schedule~\cite{Binosi2024}.

As discussed in Sec.~\ref{sec:freqBinTqst}, determining the off-diagonal terms requires projections onto states beyond those of the computational basis. Following \cref{eq:probBessel}, the frequency-bin mixing efficiency is controlled by the modulation index $\beta$, which is proportional to the ratio of the peak RF voltage swing $V_M$ and modulator half-wave voltage $V_\pi$: $\beta=\frac{\pi V_M}{V_\pi}$. Our EOPM possesses a maximum RF power limit of 30~dBm with $V_\pi=3.5$~V at 25~GHz; together these constrain the experimentally obtainable modulation index to $\beta\leq 8.97$. Specifically, for each $ij$ pairing, we select $\beta$ such that $|J_0(\beta)| \approx |J_{j-i}(\beta)|$ subject to this constraint, giving $\beta\in\{1.43, 3.83, 7.53, 4.42, 4.81, 5.11, 7.42, 7.83, 8.14\}$. As we discuss further in \cref{sec:discussion}, these choices are suboptimal in terms of statistical efficiency, yet they highlight the flexibility of tQST to succeed under a variety of experimental conditions.

Integrating the coincidences over 30~s per projection for all $d(d-1)$ superpositions at each $d$, we perform Bayesian inference on the complete datasets and obtain $N=1024$ samples $\{\bx^n\}_{n\in[N]}$ of the density matrix parameters, from which the mean and standard deviation of any quantity of interest can be computed; see Appendix~\ref{sec:mcmc} for details on the parametrization and MCMC process.

\begin{figure*}[t!]
    \centering
    \includegraphics[width=\textwidth]{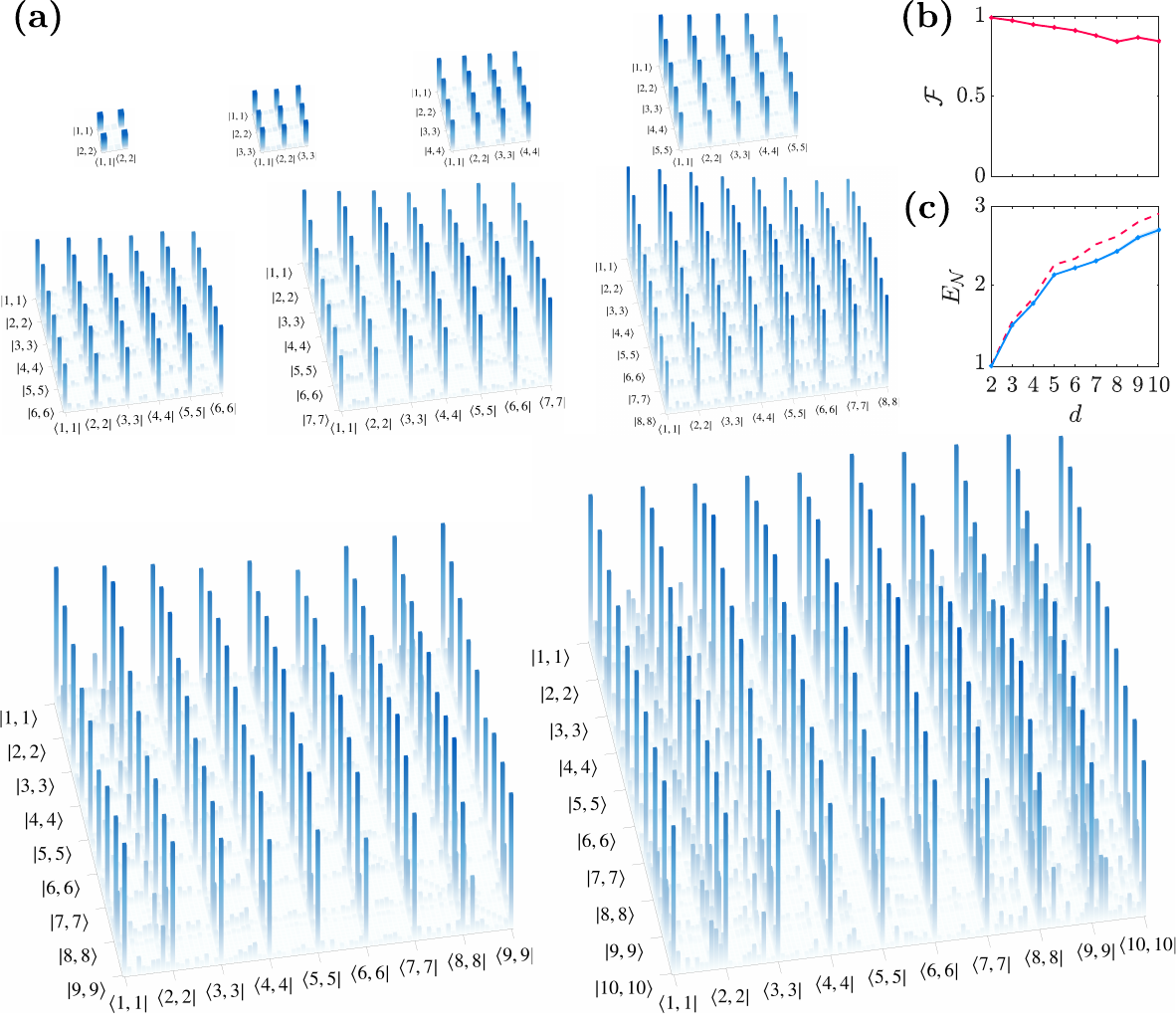}
    \caption{Experimental results. (a) Magnitude of Bayesian-estimated two-qudit density matrices with single-qudit dimension $d\in\{2,3,\ldots,10\}$. 
    (b) Fidelity with respect to Werner states with $\lambda\in\{0.982, 0.966, 0.883, 0.955, 0.836, 0.815, 0.764, 0.771, 0.747 \}$. (c) Logarithmic negativity $E_{\mathcal{N}}$ as a function of $d$ (solid blue), with the dashed red line indicating the values for the hypothesized $d$-dimensional Werner states. In both (b) and (c), shaded regions included one standard deviation in each direction.}
    \label{fig4}
\end{figure*}

The magnitude of the Bayesian mean estimates $\hrho=\frac{1}{N}\sum_{n=1}^N\rho(\bx^n)$ are plotted in \cref{fig4}(a), for qudit dimension $d\in\{2,..,10\}$. In all cases, the observed pattern confirms off-diagonal coherence and entanglement as predicted.
Beyond the main off-diagonal peaks $\rho_{(ii)(jj)}$, subsidiary peaks in a ``waffle-iron'' pattern are also evident for points of the form $\rho_{(ij)(kk)}$. Not directly measured in the thresholding procedure, the larger amplitude of these points compared to the valleys in between can be attributed to their higher Cauchy-Schwarz ceilings from the correlated $\rho_{(kk)(kk)}$ component in the product $\sqrt{\rho_{(ij)(ij)}\rho_{(kk)(kk)}}$. By contrast, in the valleys both relevant diagonal terms in $\sqrt{\rho_{(ij)(ij)}\rho_{(k\ell)(k\ell)}}$ have low amplitude, bounding these  unknown off-diagonal elements more strongly.

Significantly, the $d=10$ two-qudit Hilbert space dimension $d^2=100$ exceeds the previous 64-dimensional frequency-bin record~\cite{Lu2022b} by 56\%---additionally significant in that the number of independent parameters for a general 100-dimensional state ($10^4-1$) is 2.44$\times$ that of a 64-dimensional one. Obtained from only 190 projections, this highest-dimensional result requires less than 2\% of the $\mathcal{O}(10^4)$ measurements for traditional QST  and less than 7\% of the number of projections using the randomized approach of Ref.~\cite{Lu2022b} (3000 assuming scaling trends would have continued).

We compare each matrix sample $\rho(\bx^n)$ against a reference $\rho_\text{ref}$  via the fidelity
\begin{equation}
\label{eq:fid}
\cF^n = \left(\Tr\sqrt{\sqrt{\rho_\text{ref}}|\rho(\bx^n)|\sqrt{\rho_\text{ref}}}\right)^2,
\end{equation}
from which the mean and standard deviation over all samples $N$ gives the estimate $\hcF$ and its uncertainty, respectively.
Since dispersion has no impact on the state's entanglement quality, we compensate it numerically in the above, taking the zero-phase version of each sample through the absolute value $|\rho(\bx^n)|$.
Previous experience~\cite{Imany2019, Lu2022b} suggests that the BFC is well approximated by a Werner state~\cite{Werner1989},
\begin{equation}
\label{eq:Werner}
\rho_\text{W}=\lambda \rho_0+\frac{1-\lambda}{d^2}\mathbb{I}_{d^2},
\end{equation}
where $\mathbb{I}_{d^2}$ is the $d^2\times d^2$ identity matrix and $\lambda$ can be estimated from the CAR according to
\begin{equation}
\mathrm{CAR}=1+\frac{\lambda d}{1-\lambda}.
\end{equation}

The Bayesian-estimated fidelity with respect to the hypothesized Werner state [$\rho_\text{ref}=\rho_W$ in \cref{eq:fid}] is shown by the solid red curve in \cref{fig4}(b), with the barely visible shaded region indicating one standard deviation. The fidelity exceeds $\hat{\cF}\geq0.844$ for all qudit dimensions $d$, further validating both the performance of tQST and the suitability of the Werner-state hypothesis for describing the experimentally generated BFC state. In particular, fidelity to the hypothesized state provides a more appropriate gauge of the quality of the reconstructed BFC state than fidelity to the ideal maximally entangled state, since the former accounts for the experimentally observed noise and imperfections. For comparison, the fidelity with respect to this ideal maximally entangled target [i.e., $\rho_\text{ref}=\frac{1}{d}\sum_{i,j\in[d]}\ket{ii}\bra{jj}$ in \cref{eq:fid}] ranges from $\hcF=0.979(6)$ at $d=2$ to $\hcF=0.65(1)$ at $d=10$. 

As another measure state quality, we consider the logarithmic negativity $E_\cN$~\cite{Peres1996, Vidal2002}. An upper bound on distillable entanglement, $E_\cN=\log_2 d$ for a maximally entangled two-qudit state.  The blue curve in \cref{fig4}(c) shows the Bayesian-estimated values for all nine states, indicating genuine high-dimensional entanglement. For example, the $d=10$ case reaches $E_\cN=2.70(3)$~ebits, a number only possible for qudit dimension $d\geq 7$, confirming high-dimensional entanglement. The dashed red curve shows the values of $E_\cN$ associated with Werner states $\rho_\text{W}$ [\cref{eq:Werner}] at the experimentally measured CAR values $\text{CAR}\in\{108, 87, 77, 106, 92, 31, 27, 31, 31\}$, showing good agreement between the hypothesis and Bayesian estimates. We note that the variation in CAR is due to measurements being performed at different pump powers: we increased power at higher $d$ to reduce statistical noise under the lower projection efficiencies.

\section{Discussion}
\label{sec:discussion}
Overall, the number of projective measurements is appreciably less than both tomographically complete QST and prior randomized approaches. For example, the $d=8$ case here (corresponding to the previous frequency-bin record) has 4095 independent real parameters and was characterized in Ref.~\cite{Lu2022b} with 1920 distinct experimental settings; by contrast, we reconstruct the state using only 120 projections ($d^2$ JSI and $d^2 - d$ off-diagonal measurements). 

\begin{figure}[!t]
    \centering
    \includegraphics[width=0.40\textwidth]{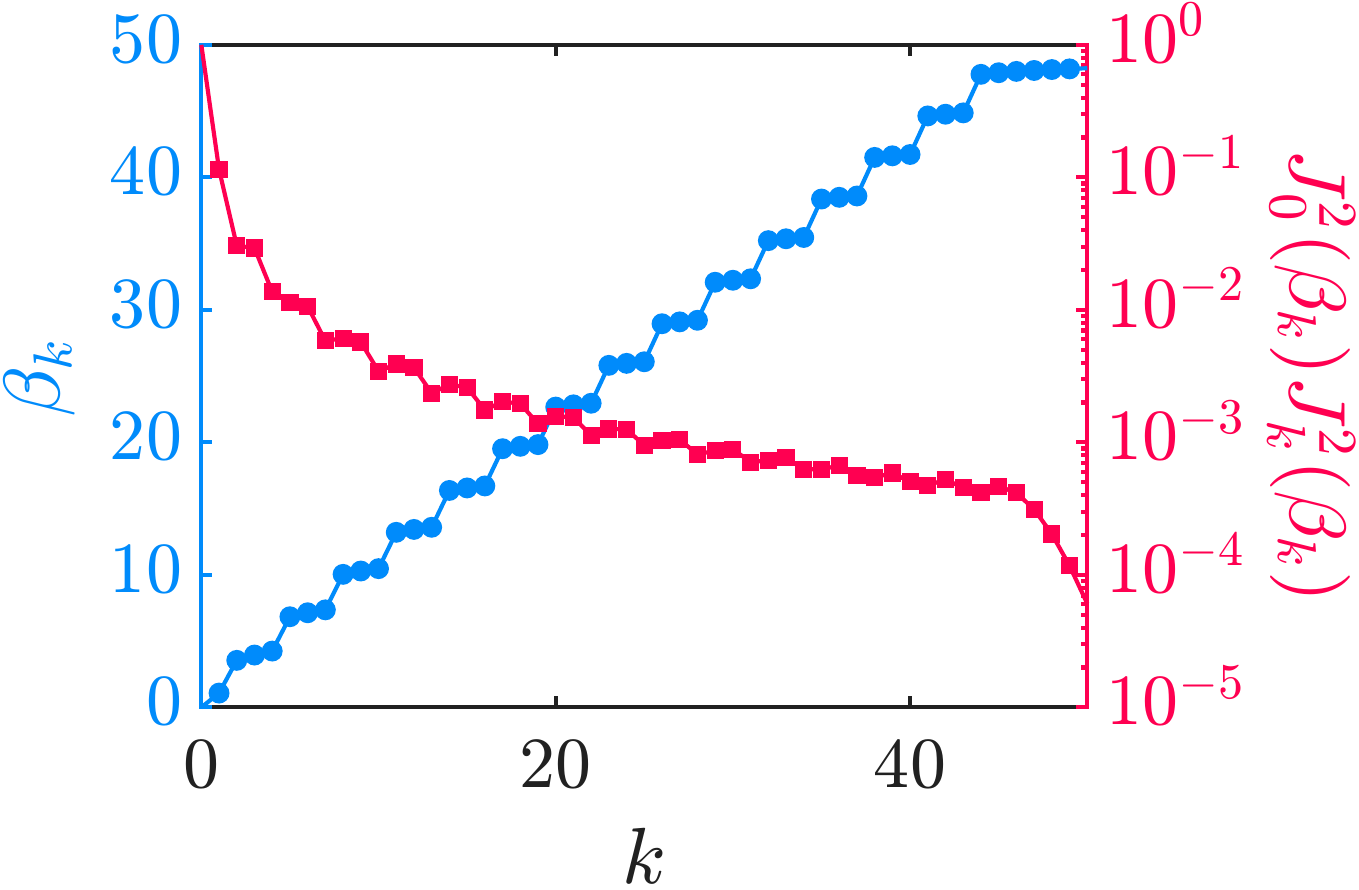}
    \caption{Optimal modulation for frequency-bin mixing. The optimal modulation index $\beta_k$ (left axis) and corresponding maximum mixing efficiency $J_0^2(\beta_k)J_k^2(\beta_k)$ (right axis) are shown as a function of frequency-bin separation $k$.}
    \label{fig:opt}
\end{figure}

Perhaps the greatest experimental challenge in reaching even higher dimensions stems from the electro-optic modulation efficiency. As highlighted in \cref{eq:probBessel}, sensitivity of the coincidence probability to the off-diagonal matrix element $\rho_{(ii)(jj)}$ is controlled by the product $J_0^2(\beta)J^2_k(\beta)$ where $k\equiv j-i$. The unitarity of electro-optic phase modulation, combined with the intrinsic spreading into sidebands beyond those of interest, ultimately limits this sensitivity.

Quantitatively, the best-case scenario obtains when the modulation index $\beta_k$ for measurement of an off-diagonal element with frequency-bin spacing $k$ is chosen according to the optimization
\begin{equation}
\label{eq:betak}
\beta_k = \argmax_{\beta\in\mathbb{R}} J_0^2(\beta)J_k^2(\beta).
\end{equation}
\Cref{fig:opt} plots these optimal indices up to $k=49$. $\beta_k$ scales linearly with $k$ with near-unit slope, confirming the intuitive relationship $\beta_k\sim\cO(k)$ noted previously for efficient electro-optic-based frequency-bin tomography~\cite{Lu2022b}, while the stair steps can be attributed to avoidance of the zeros of $J_0(\beta)$, which suppress the product $J_0^2(\beta)J_k^2(\beta)$ even when $J_k^2(\beta)$ is large. Plotted alongside these optimal indices are the corresponding maxima $J_0^2(\beta_k)J^2_k(\beta_k)$, which unsurprisingly trend downward with $k$: the more sidebands produced through electro-optic modulation, the less energy is concentrated in the interference terms of interest. Nevertheless, the reduction with $k$ is relatively gentle, remaining within an order of magnitude of the $k=9$ case until a sharp drop for $k\geq 47$. These results suggest that tQST of entangled qudits up to $d\approx 45$---a 2025-dimensional Hilbert space---should be possible with reasonable mixing probability.

Unlocking such high dimensions will require modulation indices much larger than the $\beta_{\max}=8.97$ possible in the current implementation (see \cref{sec:results}). Although cascading EOPMs in series can increase $\beta$ linearly---a technique common for increasing the bandwidth of electro-optic frequency combs~\cite{Metcalf2013, TorresCompany2014}---a more compact (and lower-loss) direction would be to pursue EOPMs with half-wave voltages significantly smaller than the $V_\pi=3.5$~V of the current device. For example, commercially available thin-film lithium niobate EOPMs with $V_\pi=0.7$~V~\cite{Hyperlight2026} would enable $\beta=44.9$ at 1~W RF input power, making ($d^2=45^2$)-dimensional two-qudit tQST in principle feasible.

Interestingly, tQST's focus on the coherences between two computational-basis states at a time suggests practical opportunities for other high-dimensional encodings as well~\cite{Erhard2020, https://doi.org/10.48550/arxiv.2604.06528}. For example, in time bins, binary two-bin measurements can be realized with a single tunable interferometer~\cite{Nowierski2016} or optical buffer~\cite{Compagnini2026}, rather than a nested tree for $d$-ary superpositions~\cite{Ikuta2017}.
In frequency encoding as well, this binary arrangement should open pathways for nonlinear spectral interference modalities in $\chi^{(2)}$~\cite{Kobayashi2016, Kobayashi2017} and $\chi^{(3)}$~\cite{Clemmen2016, Oliver2025} media. While straightforward high-dimensional mixing forms one of the key advantages of EOPM-based frequency-bin interference~\cite{Lu2022a,Lu2022b,Lu2023b}, the preceding discussion shows how most of this high-dimensional energy is lost in the tQST process. By contrast, three- and four-wave mixing processes are inherently bin-selective, enabling projections onto the states in \cref{eq:projStates} with in principle 100\% efficiency. Provided the pump fields can be tuned while retaining phase-matching, it should then be possible to perform frequency-bin tQST measurements without a dimension-efficiency tradeoff like that in \cref{fig:opt}.

Barring an unforeseen discovery exponentially reducing wall-clock times in Bayesian quantum state estimation, it is unlikely that full Bayesian inference will be able to keep pace with the $\sim$1000-dimensional Hilbert spaces envisioned for future tQST experiments, in which case maximum likelihood estimation~\cite{James2001, Lvovsky2004} or projected least-squares~\cite{Smolin2012} could take the helm as positivity-preserving alternatives.
Yet ultimately, this computational ``problem'' is an opportunity: BFC datasets beyond what can be analyzed by Bayesian techniques would herald a new era for the scaling of frequency-based photonic quantum information, with tQST serving as the key tomographic window.

\begin{acknowledgments}
We thank N.~T. Otterstrom for discussions. Funding was provided by the U.S. Department of Energy (DE-SC0026396) and the Laboratory Directed Research and Development program of Sandia National Laboratories (EPIQ, APT).
\end{acknowledgments}
\nocite{*}

\appendix
\section{MCMC details}
\label{sec:mcmc}
Let \(\rho(\mathbf{x})\) denote the density matrix in \cref{eq:rho} parameterized by some vector \(\mathbf{x}\). Bayes' theorem takes the form
\begin{equation}
\label{eq:Bayes}
\pi(\mathbf{x}) = \frac{1}{\mathcal{Z}} L_{\mathcal{D}}(\mathbf{x}) \pi_0(\mathbf{x}),
\end{equation}
where \(\pi(\mathbf{x}) = \Pr(\mathbf{x}|\mathcal{D})\) is the posterior distribution, \(L_{\mathcal{D}}(\mathbf{x}) \propto \Pr(\mathcal{D}|\mathbf{x})\) is the likelihood of the observed data \(\mathcal{D}\), \(\pi_0(\mathbf{x}) = \Pr(\mathbf{x})\) is the prior, and \(\mathcal{Z} = \int d\mathbf{x}\, L_{\mathcal{D}}(\mathbf{x})\pi_0(\mathbf{x})\) is the normalization constant.

\begin{figure*}[t!]
  \centering
  \includegraphics[width=1\textwidth]{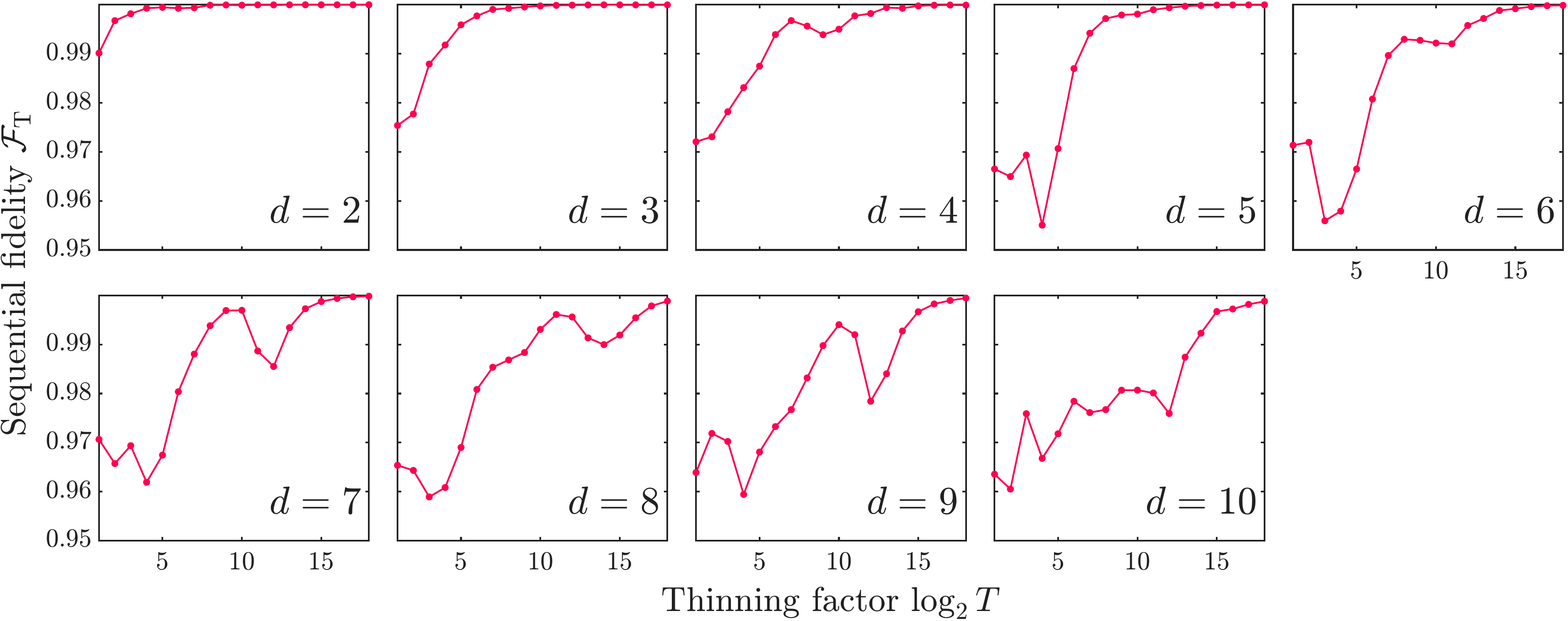}
\caption{Convergence analysis of Bayesian quantum state tomography  for Bayesian quantum state tomography two-qudit states $d\in\{2,3,\ldots,10\}$, quantified via the sequential fidelity $\cF_T$ between Bayesian mean $\hrho_T$ at thinning $T$ and its previous value $\hrho_{T/2}$ at thinning $T/2$.}
\label{fig5}
\end{figure*}

We follow previous work~\cite{Lu2022b, Alshowkan2022c, Lu2023b} and take a Bures prior distribution~\cite{Sommers2003, Osipov2010} with a normally distributed flux $K(\bx)=K_0(1+0.1x_{4d^4+1})$, where $K_0$ is equal to the total number of observed coincidences in the computational basis. Under this parametrization, $\rho(\bx)$ is constructed by a $d^2\times d^2$ complex matrix $G(\bx)$ (defined by $2d^2$ real numbers) and a $d^2\times d^2$ unitary marix $U(\bx)$ (constructed by the Mezzadri algorithm~\cite{Mezzadri2007} from $2d^2$ real inputs). Accordingly, we can write
\begin{equation}
\rho(\bx)=\frac{W(\bx)W^\dagger(\bx)}{\Tr W(\bx)W^\dagger(\bx)} \; ; \; W(\bx)=[\mathbb{I}_{d^2}+U(\bx)]G(\bx).
\end{equation}

This model relies on a length-($4d^4+1$) parameter vector $\bx=(x_1,...,x_{4d^4+1})$ with a standard normal prior $\pi_0(\bx)=\prod_{k=1}^{4d^4+1}\frac{1}{\sqrt{2\pi}}e^{-\frac{1}{2}x_k^2}$ and likelihood
\begin{equation}
\label{eq:likelihood}
L_\cD(\bx)=\prod_{s=1}^S e^{-K(\bx)p_s(\bx)}[K(\bx)p_s(\bx)]^{N_s},
\end{equation}
for $\bcD=\{N_s\}_{s\in[S]}$ coincidences observed for each projection $s\in[S]$. In our case, $S=d^2+d(d-1)$ from $d^2$ computational basis measurements with probabilities $p_s(\bx)=\rho_{(ij)(ij)}$ and $d(d-1)$ off-diagonal projection probabilities $p_s(\bx)=p_{ij}$ [\cref{eq:probBessel}] for $1\leq i < j \leq d$ and $\theta\in\{0,\frac{\pi}{2}\}$.

Computationally, we sample from \cref{eq:Bayes} via a parallel preconditioned Crank-Nicolson (pCN) algorithm~\cite{Cotter2013,Lukens_2020,Nguyen2025}. Assuming a chain length of $NT$, the true posterior $\pi(\bx)$ can be approximated by the distribution
\begin{equation}
\label{eq:BayesHat}
\hpi(\bx) = \frac{1}{N}\sum_{n=1}^N \delta(\bx-\bx^n),
\end{equation}
i.e., a superposition of point masses at the $N$ saved samples, such that the $n$-th saved sample $\bx^n$ corresponds to the $nT$-th step in the full MCMC chain. The estimated Bayesian mean $\hrho$ follows as
\begin{equation}
\label{eq:BME}
\hrho = \int d\bx\,\hpi(\bx)\rho(\bx)
= \frac{1}{N}\sum_{n=1}^N \rho(\bx^n),
\end{equation}
where, for all results reported in the main text, we take $(N,T)=(2^{10},2^{18})$.

Ensuring adequate convergence for such long chains can prove challenging. A useful diagnostic is the  sample autocorrelation function (ACF), which for density matrices can be defined as~\cite{Nguyen2025}
\begin{equation}
\label{eq:ACFrho}
c[\ell] = \frac{1}{C}\sum_{n=1}^{N-\ell} \mathfrak{Re}\left\{ \Tr [\rho(\bx^n) - \hat{\rho}]^\dagger [\rho(\bx^{n+\ell}) - \hat{\rho}]\right\},
\end{equation}
where $C$ is a normalization constant such that $c[0]=1$ and $\ell$ is the lag between samples. The integrated autocorrelation time (IACT) is then
\begin{equation}
\label{eq:IACT}
\tau = 1 + 2\sum_{\ell=1}^{\ell_\mathrm{max}} c[\ell],
\end{equation}
with $\ell_\text{max}$ some maximum lag chosen beyond the point at which the ACF converges to approximately zero. For an ideal set of uncorrelated samples, $c[\ell]=\delta[\ell]$ and $\tau=1$; in practice, the closeness of $\tau$ to unity provides a measure of the convergence.
In the current context, we use the IACT to rank the quality of independently run MCMC chains and select that with an IACT closest to unity for the final sample set $\{\bx^n\}_{n\in[N]}$. (Because of wide variability in convergence per chain, we opted for this approach rather than pooling all chains together as in Ref.~\cite{Nguyen2025}.)
Empirically, our saved chains obtain $\tau \in\{0.8, 0.9, 1.1, 1.0, 1.1, 1.6, 14.4, 4.9, 8.5\}$ for $d \in\{2,3,4,5,6,7,8,9,10\}$. 
The general increase in $\tau$ at higher dimensions stems from the inherent difficulty of exploring larger parameters spaces, where MCMC naturally requires more steps to fully decorrelate.

While the IACT values certainly rank the utility of competing chains, they do not directly bound many quantities of interest, such as the fidelity of the estimated quantum state $\hrho$ to the true Bayesian mean obtained in the limit $NT\rightarrow\infty$. In order to confirm the sufficiency of the ($NT=2^{28}$)-long chains adopted here, we additionally compute the sequential fidelity $\cF_T$---namely, the overlap between the estimate $\hrho_T$ at thinning $T$ and its value $\hrho_{T/2}$ at thinning $T/2$: 
\begin{equation}
\label{eq:seqFid}
\cF_T=\left(\Tr\sqrt{\sqrt{\hrho_T}\hrho_{T/2}\sqrt{\hrho_T}}\right)^2.
\end{equation} \Cref{fig5} plots $\cF_T$ for all experimental datasets. $\cF_T>0.94$ throughout, even at extremely small thinning values, and all cases converge to $\cF_T>0.999$ at $T=2^{18}$, implying that all saved chains have converged to their true Bayesian means with $\sim$99.9\% fidelity.

\bibliography{references}
\end{document}